\documentclass[preprint]{elsarticle}
\usepackage{color}
\def\Red#1{{\color{black} #1}}
\title{An alternative observable to estimate $k_{\rm eff}$ in fast ADS.}

\author[infn,unige]{D. Chersola}
\ead{xxx}

\author[infn,centrofermi]{G. Ricco\corref{cor1}}
\ead{ricco@ge.infn.it}

\author[infn,centrofermi]{M. Ripani}
\ead{marco.ripani@ge.infn.it}

\author[infn]{P. Saracco}
\ead{paolo.saracco@ge.infn.it}

\cortext[cor1]{Corresponding author}

\fntext[infn]{I.N.F.N. - Sezione di Genova, V. Dodecaneso, 33 - 16146 Genova (Italy)}
\fntext[unige]{DIME/TEC - University of Genova, V. all'Opera Pia, 15/A - 16145 Genova (Italy)}
\fntext[centrofermi]{CENTRO FERMI - Compendio del Viminale  -€"  Piazza del Viminale 1 -€" 00184 Rome (Italy)}

\begin{document}

\begin{abstract}

In this paper we examine an alternative method for determining the neutron multiplication factor $k_{\rm eff}$ in fast  subcritical reactors, based on the observable obtained by calculating the ratio between the ``fast`` component of the neutron flux (in this case by fast we mean neutrons with kinetic energy above 0.5 MeV) and the whole flux at all energies.
The assumption behind the present study was that the above mentioned observable would be only sensitive to $k_{\rm eff}$, provided a peripheral fuel rod is considered (i.e. far from the neutron source in subcritical systems) and $k_{\rm eff}$ is sufficiently high.
Indeed, our results show that, for $k_{\rm eff}>$ 0.7, the ratio calculated for \Red{LFR or GFR models} turns out to show a quite similar trend well approximated by a straight line, while a different slope is observed for a sodium reactor.
In our opinion, the results obtained show that such observable, after appropriate calibrations on the real systems under consideration, may be used for determining $k_{\rm eff}$ in all operational conditions, in particular for subcritical systems with different types of \Red{neutron }source and different values of the beam current.  

\end{abstract}


\maketitle

\section{Introduction}
The neutron multiplication factor $k_{\rm eff}$ is not only a fundamental quantity for the understanding of the physical behaviour of fast subcritical reactors (ADS), but also an essential parameter to monitor the subcriticality level as required for safe operations: indeed not only the maximum available power, but also the reactor kinetics and dynamics \citep{SarRic} depend on it. However, $k_{\rm eff}$, or equivalently the reactivity, is not directly an observable and its determination is essentially an inverse problem, whose solution generally relies on indirect techniques, so that appropriate methods have to be developed for its experimental determination.

\Red{Many theoretical methods have been deeloped in the past to relate $k_{\rm eff}$ to physically measurable quantities: they are based on generalized perturbation theory \cite{Bruna}, on source multiplication \cite{Green} or on point kinetics model \cite{Chevret}-\cite{Mila}, whose validity is questionable far from criticality. The crucial point to observe is that in an ADS it is necessary to prevent neutron flux divergence both in normal operating conditions and in abnormal accident condition: even assuming $k_{\rm eff}$ as a gauge for criticality, the concept of "margin to criticality" is questionable \cite{SarMar}, so that monitoring the $k_{\rm eff}$ value during normal operations is a required safety condition for non zero power facilities \cite{MUSE}. European FP7 project FREYA was devoted essentially to this task. \cite{UyttPhysor}}

These different models have been tested against experimental results from MUSE \cite{MUSE}, YALINA\cite{Yalina}, DELPHI \cite{Delphi} and, more recently, VENUS-F\cite{Venus}: results differ from experiment to experiment and depend on the subcriticality level. During normal operation an ADS will run in continuous mode, which means that the accelerated beam is uninterrupted and the  \Red{thermal} power-to-\Red{accelerator-}current indicator provides an on-line measurement of the reactivity. However, the proportionality constant in the power-to-current indicator has to be regularly checked by repeatedly applying absolute reactivity monitoring techniques, which require the accelerator to deliver specific beam time structures \cite{Carta}. It is worth noticing that the power-to-current indicator will depend on the type of neutron source - \Red{because different target configurations and/or beam energies produce different neutron yields and spectra} - and its value relies on the correct measurement of two independent quantities.

A major difficulty in applying and interpreting such techniques is the simultaneous presence during the transients of many of the normal modes of the system, entailing the non separability of the neutron flux into space and time dependence: therefore, time-dependent local flux measurements will yield different $k_{\rm eff}$ estimates that require to be spatially corrected, no matter which specific technique is employed, like e.g. source-jerk \cite{Chevret}, $k_p$ method\cite{Lecolley}, area method\cite{Berglof,Becares,Mila}, etc. 

\Red{Our study concerns subcritical fast systems, where the core does not contain any moderator material. We considered three cases, where the fuel pins are embedded in lead, gas, or sodium, with reference to the fast reactor types considered in the so-called Generation IV framework. Since fission neutrons scatter mainly on lead, low-density helium gas, or sodium (besides scattering on the fuel itself, where some limited moderation may occur on oxygen atoms contained in the compounds), the resulting neutron energy spectrum is hard or, generally speaking, "fast". In order to define more precisely the "fast" component of the spectrum, we distinguish between neutron with kinetic energy below 0.5 MeV and above 0.5 MeV. Although somewhat arbitrary, the choice of this threshold corresponds to the minimum energy for fission to occur in some Minor Actinides (MA: neptunium-247, americium-241,...), so it has a physical meaning in that "fast" neutrons will be able to produce a signal in fission chambers coated with MA, while "slow" neutrons will not. }

An alternative observable, possibly related to $k_{\rm eff}$, is given in ADS by the measurement of the relative fast flux component on peripheral fuel rods.  If we assume that the primary fast neutrons from the source are almost completely absorbed in the innermost part of the core, the fast component of the flux in the peripheral rods should be mostly due to those secondary fission neutrons, which are only partially slowed down by interactions with fuel and coolant. Based on this assumption, we defined a type of spectral index as a measure of the relative fast flux component, by introducing the ratio R 
\begin{equation}
R= \frac{ \int_{E_{th}}^{}{\Phi(E)dE } } { \int{\Phi(E)dE } }
\label{eqn:ratio}
\end{equation} 
where $\Phi$
is the neutron flux averaged over the whole rod, E is the neutron energy and  $E_{th}$ is a specific energy threshold that in the rest of the paper will be assumed equal to 0.5 MeV. In practice  $E_{th}$ should be a physics, hardware or software threshold set in the device used for the measurement. In our assumption, R should be essentially determined, for peripheral rods and a specific detector threshold $E_{th}$, by the fission rate and, therefore, by a function of the neutron multiplication factor $k_{\rm eff}$. If that were the case, such an observable would offer the advantage of being measurable in all operational conditions. Moreover, it should be relatively independent from the type of neutron source, being essentially a characteristic of the specific core, and would be derived from two quantities measured with the same device. In particular, it could be used for monitoring the reactivity level periodically when the accelerator is running in continuous mode, thereby providing a real-time, source- and current- independent cross check of the power-to-current indicator without turning off the beam.

However, the extraction of $k_{\rm eff}$ from the direct measurement of R may be affected and complicated by the dependence on other parameters such as geometry (rod dimensions, fuel to coolant volume ratio etc.) or composition (fuel and coolant materials).
In order to investigate to some extent such a dependence, we performed simulations of R$(k_{\rm eff})$ using MCNPX and different fast ADS reactor models. The neutron source in all simulations, except the comparison in Table 1, was a 70 MeV proton beam impinging on a 5 cm thick and 4 cm radius  ${^9}$Be target, producing a fast neutron spectrum accurately measured in ref. \citep{OsiEtAl1,OsiEtAl2}. 
\Red{Clearly, the R value may be slightly dependent on local features of the core - like, e.g. the measurement position or the insertion of impurities of different materials - so that for executive projects the stability (robustness) of the method should be carefully studied in order to  calibrate the relationship between R and $k_{\rm eff}$.}

\section{The model}
The reactor models analyzed in this paper all assume the same cylindrical boundaries: an inner core with 50 cm radius and 90 cm height, surrounded by a 120 cm radius and 160 cm high lead reflector, in turn contained within a 5 cm thick AISI steel vessel. The system features a 6 cm radius central axial beam pipe hosting the Beryllium target. The pipe is filled with helium gas for the purpose of target cooling.  For the core, four different configurations were considered:

\begin{enumerate}
\item A lead type fast reactor core (which we will call reference model)  designed as a 1.6x1.6x90 $cm^3$ hexahedral type lattice containing the 0.37 cm radius fuel rods surrounded by a 0.05 cm thick AISI steel cladding and embedded in solid lead. The fuel was a mixture of ${}^{238}U$ and ${}^{235}U$: $k_{\rm eff}$ was varied by increasing up to 30\% the ${}^{235}U$ enrichment. 

\item A lead type fast reactor core designed as a 2.3x2.3x90 $cm^3$ hexahedral type lattice containing 1 cm radius fuel rods   surrounded by a 0.1 cm thick steel cladding and embedded in lead. The fuel composition was as in 1) with relative $^{235}$U enrichment varied up to 12\%.

\item A fast waste burner core designed as a 1.14x1.14x90 cm$^3$ hexahedral lattice filled with 0.5 cm \Red{radius} rods  surrounded by a 0.05 cm thick steel cladding and embedded in He gas. The fuel, a futurix6 \citep{futurix} type of ceramic actinide mixture, could be varied again by changing the Pu isotopes mass percentage. In this case, we found that \Red{no material can be interposed} between the fuel elements to obtain a relatively high value of $k_{\rm eff}$ which is more of interest in this context.

\item A sodium type fast reactor core with the same geometry as (1), with liquid sodium replacing lead and using MOX \citep{MOX} as fuel: $k_{\rm eff}$ was varied by changing up to 30\% the Pu isotopes mass percentage in the mixture. 
\end{enumerate}

\begin{figure}[!h]
\centering
\includegraphics[width=12cm]{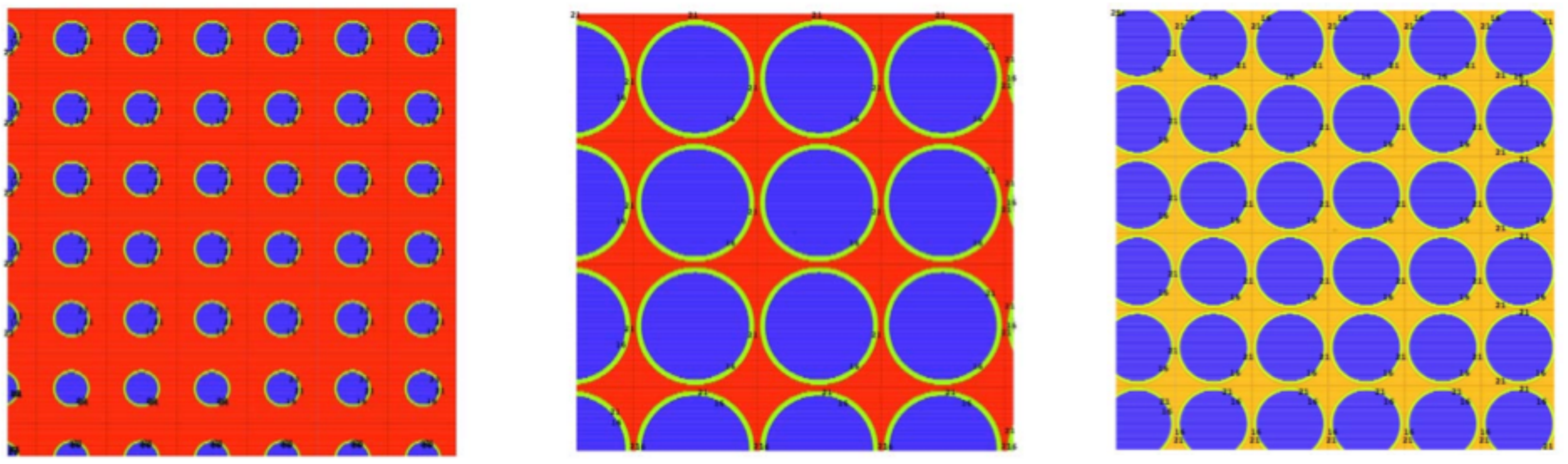}
\caption{Geometries of the models studied. Left: geometry for models (1) and (4); center: model (2); right: model (3).}
\label{Fig:class}
\end{figure}

In every case the average neutron energy spectrum in a peripheral rod at minimal distance from the core boundary was computed as a MCNPX tally and R evaluated according to Eqn.~\ref{eqn:ratio}.

\begin{table}[h]
\begin{center}
\begin{tabular}{|r|r|r|r|r|}\hline
$k_{\rm eff}$ 	& 	Flux  		& Total Flux  		& R			& 	error on R		\\
       			&  $E_n > 0.5$ MeV	& model (1)    	& model (1)		& 			\\ 
       			&  model (1)	   	& 70 MeV   		& 70 MeV		& 			\\ 
       			&   70 MeV  		& 		   	& 			& 			\\   \hline
0.0950		& 	3.035e-05	& 3.295e-05		& 0.0787		& 0.0003 		\\
0.2276		& 	3.531e-05	& 3.934e-05		& 0.1020		& 0.0004		\\
0.3420		& 	4.117e-05	& 4.688e-05		& 0.1216 		& 0.0004		\\
0.5187		&	5.490e-05	& 6.464e-05		& 0.1510 		& 0.0003		\\
0.5906		& 	6.354e-05	& 7.612e-05		& 0.1650		& 0.0004 		\\
0.6568		&	7.455e-05	& 9.053e-05		& 0.1765		& 0.0003		\\
0.7171		&	8.866e-05	& 1.092e-4		& 0.1880		& 0.0004	 	\\
0.7750		&	1.073e-4	& 1.336e-4		& 0.1970		& 0.0003		\\
0.8222		&	1.327e-4	& 1.675e-4		& 0.2070		& 0.0004	 	\\
0.8717		&	1.711e-4	& 2.185e-4		& 0.2170		& 0.0003		\\
0.9129		&	2.336e-4	& 3.016e-4		& 0.2250		& 0.0004	 	\\
0.9550		&	3.560e-4	& 4.646e-4		& 0.2340		& 0.0005	 	\\ \hline
\end{tabular}\end{center}
\caption{\label{table1} Neutron fluxes, in neutron/cm$^2$/sec/incident particle, for model (1) coupled to neutron source given by 70 MeV protons on Be target, together with corresponding ratio R along with its error. }
\end{table}

\begin{table}[h]
\begin{center}
\begin{tabular}{|r|r|r|r|r|}\hline
$k_{\rm eff}$	& Flux  		& Total Flux  	& R			& error on R		\\
       			&  $E_n >$ 0.5 MeV	& model (1)   & model (1)		& 			\\ 
       			&  model (1)	   	& 14 MeV   	& 14 MeV		& 			\\ 
       			&   14 MeV  		& 		& 			& 			\\ \hline 
0.0950		&  4.063e-4		& 4.440e-4	&	0.08490	& 0.00009		\\
0.2276		&  4.698e-4	 	& 5.257e-4	&	0.1060	& 0.0001		\\
0.3420		&  5.417e-4		& 5.417e-4	&	0.1213	& 0.0001		\\
0.5187		&  7.158e-4		& 8.457e-4	&	0.1535	& 0.0001		\\
0.5906		&  8.294e-4		& 9.945e-4	&	0.1660	& 0.0002		\\
0.6568		&  9.704e-4		& 1.180e-3	&	0.1780	& 0.0002		\\
0.7171		&  1.146e-3		& 1.412e-3	&	0.1880	& 0.0002		\\
0.7750		&  1.388e-3		& 1.725e-3	&	0.1980	& 0.0002		\\
0.8222		&  1.709e-3		& 2.159e-3	&	0.2080	& 0.0002		\\
0.8717		&  2.195e-3		& 2.804e-3	&	0.2170	& 0.0003		\\
0.9129		&  2.994e-3		& 3.868e-3	&	0.2260	& 0.0003		\\
0.9550		&  4.546e-3		& 5.936e-3	&	0.2340	& 0.0004		\\ \hline
\end{tabular}\end{center}
\caption{\label{table2} Neutron fluxes, in neutron/cm$^2$/sec/incident particle, for model (1) directly coupled to neutron source given by 14 MeV neutrons, together with corresponding ratio R along with its error. }
\end{table}

In order to give a numerical illustration of the results, we report them explicitly for model (1) in tabular form: Table~\ref{table1} reports the two integral fluxes used in the numerator and denominator of the ratio R, respectively, as well as the ratio R and its error for  a \Red{source obtained by a} 70 MeV proton \Red{beam hitting a beryllium target and producing neutrons with an evaporation energy spectrum similar to the spallation one, except for the kinematical limits \cite{OsiEtAl2}}.
In order to verify our hypothesis that periferal neutrons are weakly sensitive to the specific source at the core center, we performed the same simulation assuming 14 MeV monochromatic neutrons as obtained in well-known Deuteron-Tritium sources. Table~\ref{table2} reports the same quantities for the latter case.

\begin{figure}[!htb]
\centering
\includegraphics[width=12cm]{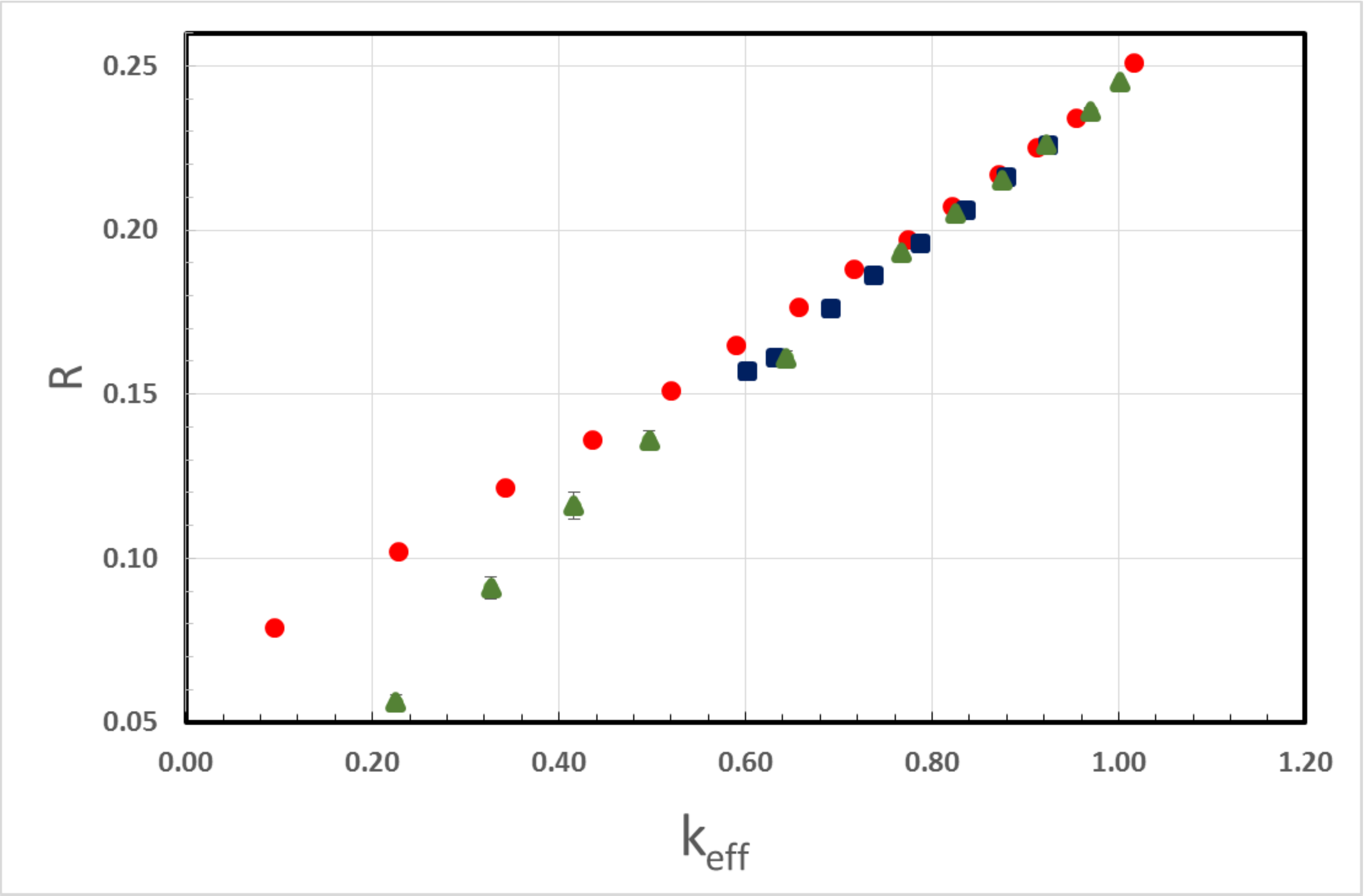}
\caption{R for a boundary rod for models (1) (red circles), (2) (green triangles) and (3) (blue squares). Error bars are smaller or equal to the marker size. \Red{Last red point for $k_{eff}=1.00xx$ was obtained with a kcode run, while all others are been obtained with fixed source runs.}}
\label{Fig:fig3}
\end{figure}

Fig.~\ref{Fig:fig3} provides a graphical representation of the results, where it can be seen that curves corresponding to models (1), (2) and (3) clearly approach each other for high $k_{\rm eff}$, converging to very similar R($k_{\rm eff} )$ functions.

\begin{table}[h]
\begin{center}
\begin{tabular}{|r|r|r|}\hline
c$_0$					& c$_1$  					& c$_2$ 				\\ \hline 
0.2425 $\pm$	0.00028 	&  - 0.201 $\pm$ 0.0015		& 0.023  $\pm$ 0.0015		\\
0.2435 $\pm$	0.00073		&  - 0.210 $\pm$ 0.0065	 	& -0.034 $\pm$ 0.0094		\\
0.242   $\pm$  0.0012		&  -0.21    $\pm$ 0.012			& -0.013  $\pm$ 0.028		\\ \hline
\end{tabular}\end{center}
\caption{\label{table3} Coefficients of quadratic fit to the ratio R for models (1), (2) and (3). }
\end{table}

\begin{table}[h]
\begin{center}
\begin{tabular}{|r|r|}\hline
c$_0$					& c$_1$  					\\ \hline 
0.2432 $\pm$ 0.00033		&  - 0.202 $\pm$ 0.0019		\\
0.2438 $\pm$ 0.00072		&  - 0.22 $\pm$ 0.0050	 		\\
0.242 $\pm$ 0.0008		&  -0.215 $\pm$ 0.0043			\\ \hline
\end{tabular}\end{center}
\caption{\label{table4} Coefficients of linear fit to the ratio R for models (1), (2) and (3), limited to $k_{\rm eff}>$ 0.7. }
\end{table}

The main features of the results for this peripheral rod can be summarized as follows:
\begin{itemize}

\item the ratio R shows a smooth, monotonically increasing dependence on $k_{\rm eff}$, which can be reproduced in the whole range, inside errors, by a simple quadratic, almost linear, function; indeed, a quadratic fit the data of the form
R($k_{\rm eff}$)=c$_0$ + c$_1$ (1-$k_{\rm eff}$) + c$_2$ (1-$k_{\rm eff}$)$^2$ 
yields the coefficients reported in Table~\ref{table3}; 
by fitting instead the data above $k_{\rm eff}$=0.7 with a linear function, we get the coefficients reported in  
Table~\ref{table4}, thereby confirming the similarity of the trends which is evident by visual inspection of the graph 
(it is worth noticing that, for model (1), the supercritical point at $k_{\rm eff}$=1.017 appears to be higher than both the quadratic and the linear fit);

\item  the average neutron spectrum has a very weak dependence from the energy distribution of the primary source neutrons; in fact, a difference is seen only for $k_{\rm eff}<0.7$, with a maximum of about 8 $\%$ for very low $k_{\rm eff}$=0.095, while there is no difference for $k_{\rm eff}>0.7$, as expected from our assumption that for sufficient multiplication fission neutrons will dominate the spectrum at the periphery of the system. 

\end{itemize}

\begin{figure}[!htb]
\centering
\includegraphics[width=12cm]{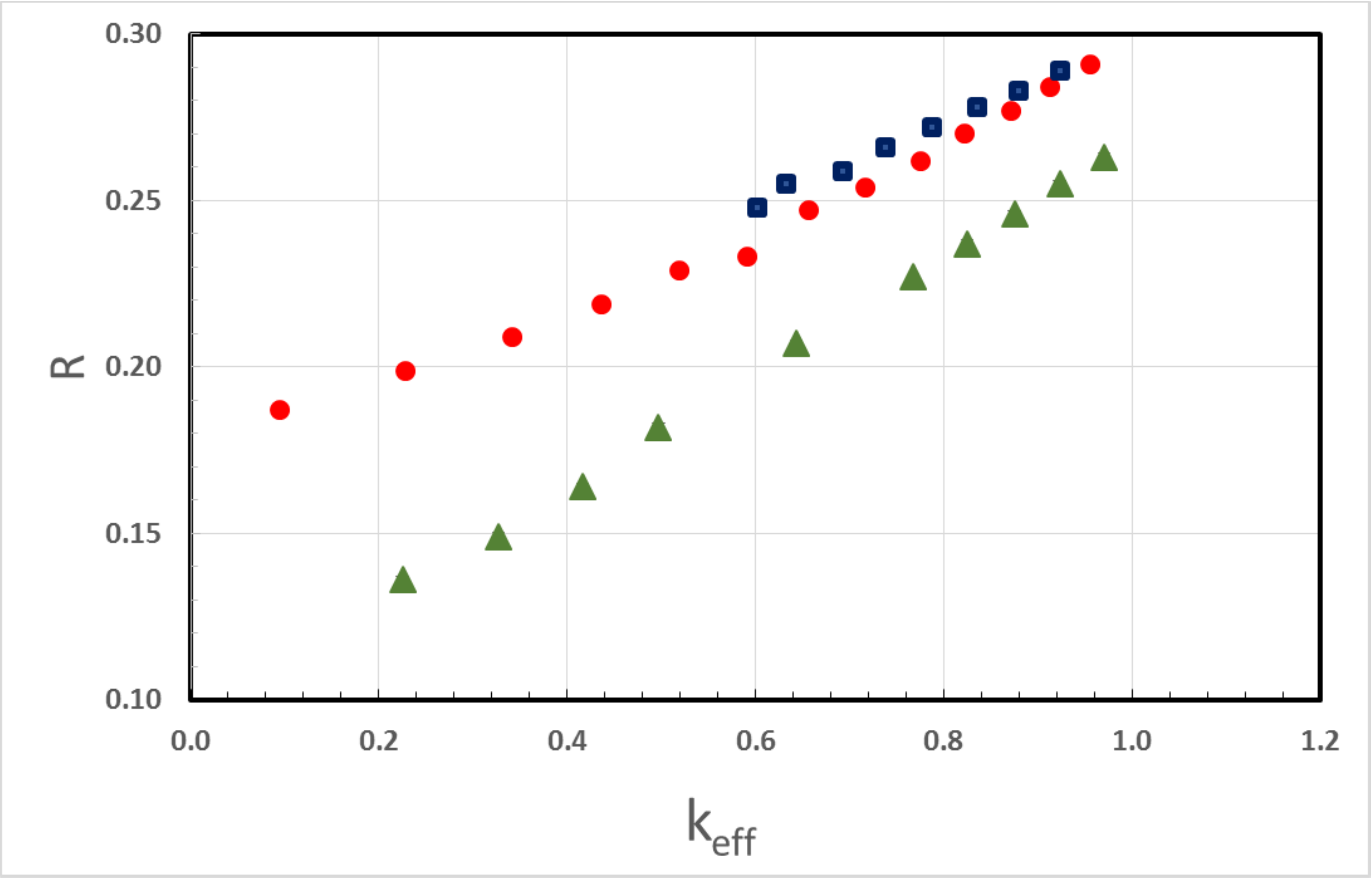}
\caption{R for an intermediate rod for models (1) (red circles), (2) (green triangles) and (3) (blue squares). Error bars are smaller or equal to the marker size.}
\label{Fig:fig2}
\end{figure}

The above results provide an indication that R($k_{\rm eff}$) is a smooth function in the whole subcritical $k_{\rm eff}$ range. As mentioned above, as a next step it is important to investigate other possible dependences from  reactor parameters, like geometry and composition. To this purpose, we also computed the R observable for an intermediate rod in each core model. The results for the intermediate rod, plotted in Fig.~\ref{Fig:fig2} for models (1), (2), and (3), clearly show in all cases a regular behaviour but a strong dependence from the assumed reactor model, even for similar lead cores. 
It is worth noticing that the ratio R for this intermediate rod appears larger than for the boundary rod, indicating a harder neutron spectrum. This is compatible with our hypothesis that, while at the boundary there is no memory of the primary source neutrons, in an intermediate position some of the more energetic source neutrons (whose kinetic energy extends up to several tens of MeV) can still be present and namely make the spectrum harder.

\Red{Then most of the flux at the boundary comes from fission neutrons that have been produced largely in the neighbourhood}. 
This interpretation is supported by the fact that the numerical convergence between the different models, as well as the source independence indicated by the comparison of Tables 1 and 2, occur for  $k_{\rm eff}>0.6 /0.7$, corresponding to a neutron multiplication 1.5/2.3, or in other words they occur when the number of fission neutrons produced is at least about twice the number of source neutrons. 

We should remark that the models used for this study are quite simplistic in terms of geometry and description of structural materials. 
For instance, the cylindrical core  boundary causes a cut off of the volume of some of the  rods intercepting the boundary itself,  possibly affecting the flux evaluation. Moreover, structural materials are reduced in the core with respect to a realistic configuration, thereby producing less absorption and moderation effects than in a real system. 
Finally, the contribution of neutrons produced in the rod itself is certainly enhanced since the flux is computed and averaged over the rod volume.

\begin{figure}[!htb]
\centering
\includegraphics[width=7cm]{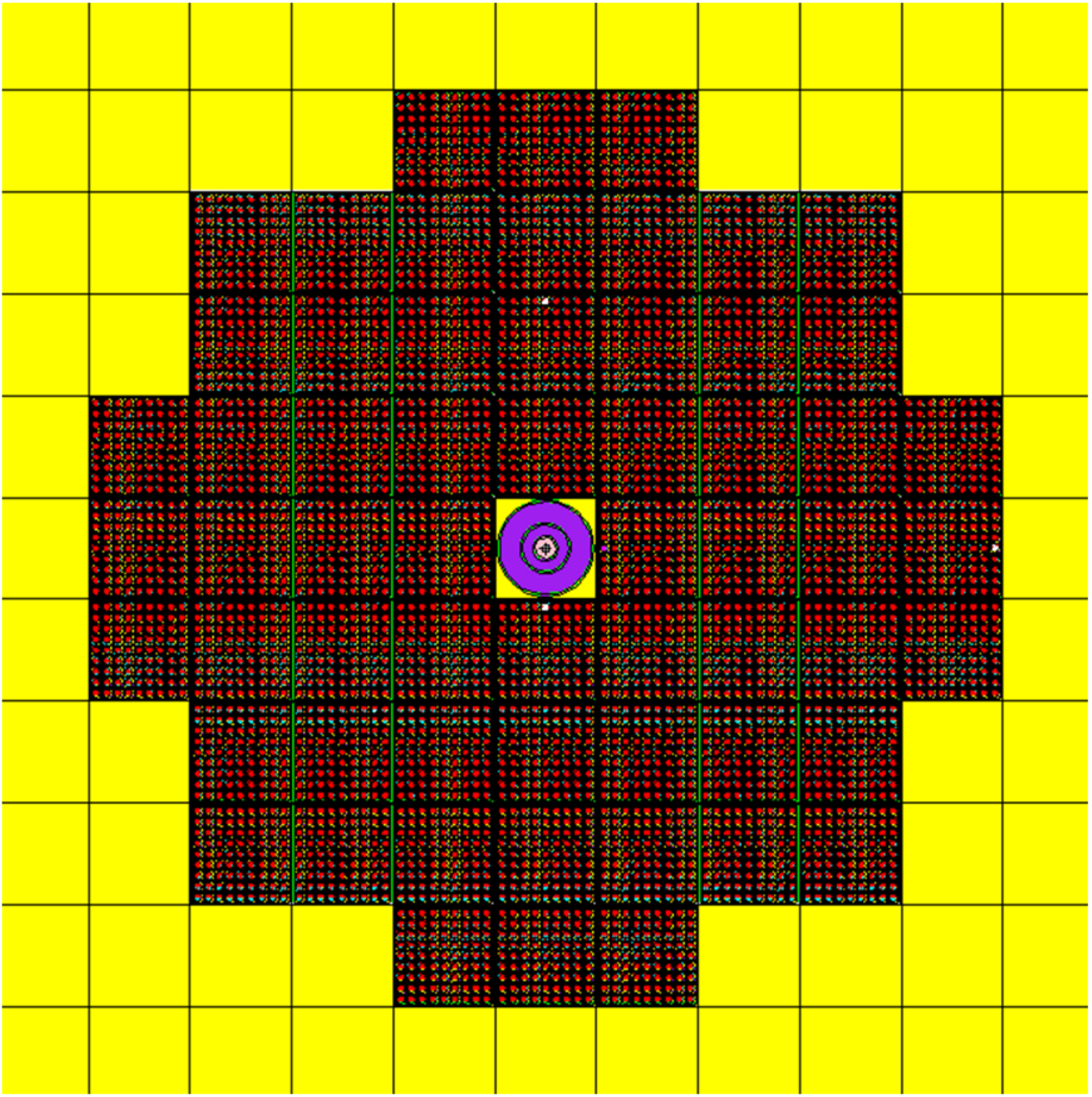}
\includegraphics[width=12cm]{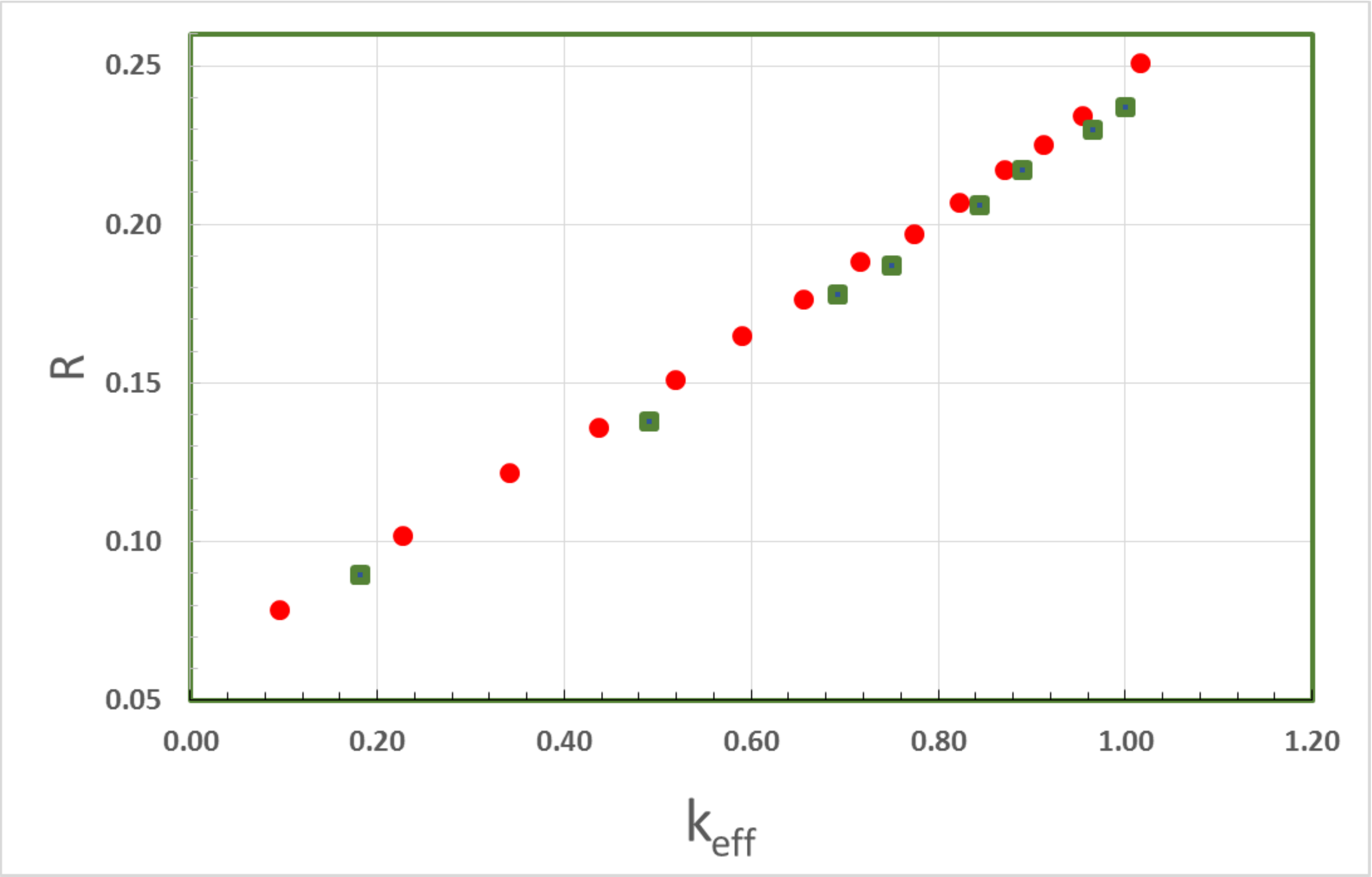}
\caption{Core shape for LEADS model (top) and corresponding R function (bottom, green squares), compared to model (1) (red circles). Error bars are smaller or equal to the marker size.}
\label{Fig:fig4a}
\end{figure}

In order to check these points we have assumed, as a more realistic model, the design of a research fast ADS, based on solid lead, described in ref. \citep{LEADS}, using again as a neutron source a 70 MeV proton beam on a $^9$Be target.
In this model, hereafter called LEADS, the fuel/lead lattice in the core is similar to model (1), but:
\begin{itemize}
\item The core shape (Fig.~\ref{Fig:fig4a}) is not any more exactly cylindrical and the boundary rods are all correctly described.
\item More structural materials, like the steel boxes containing the fuel assembly and the aluminum cooling channels for the Helium gas, are included.
\item The  fuel fills  only  the lower  half of the  test bar, while the few $cm^3$ tally volume where the flux is averaged is positioned in the empty upper half.
\end{itemize}

The simulations yield (Fig.~\ref{Fig:fig4a}, bottom graph)  an R($k_{\rm eff} )$ curve with very similar slope, but slightly lower absolute values than the previous simplified lead models: a result that, in view of the considerably higher complexity of this system, confirms in our opinion the observed convergence as a quite general feature of this type of systems. 

\begin{figure}[!htb]
\centering
\includegraphics[width=12cm]{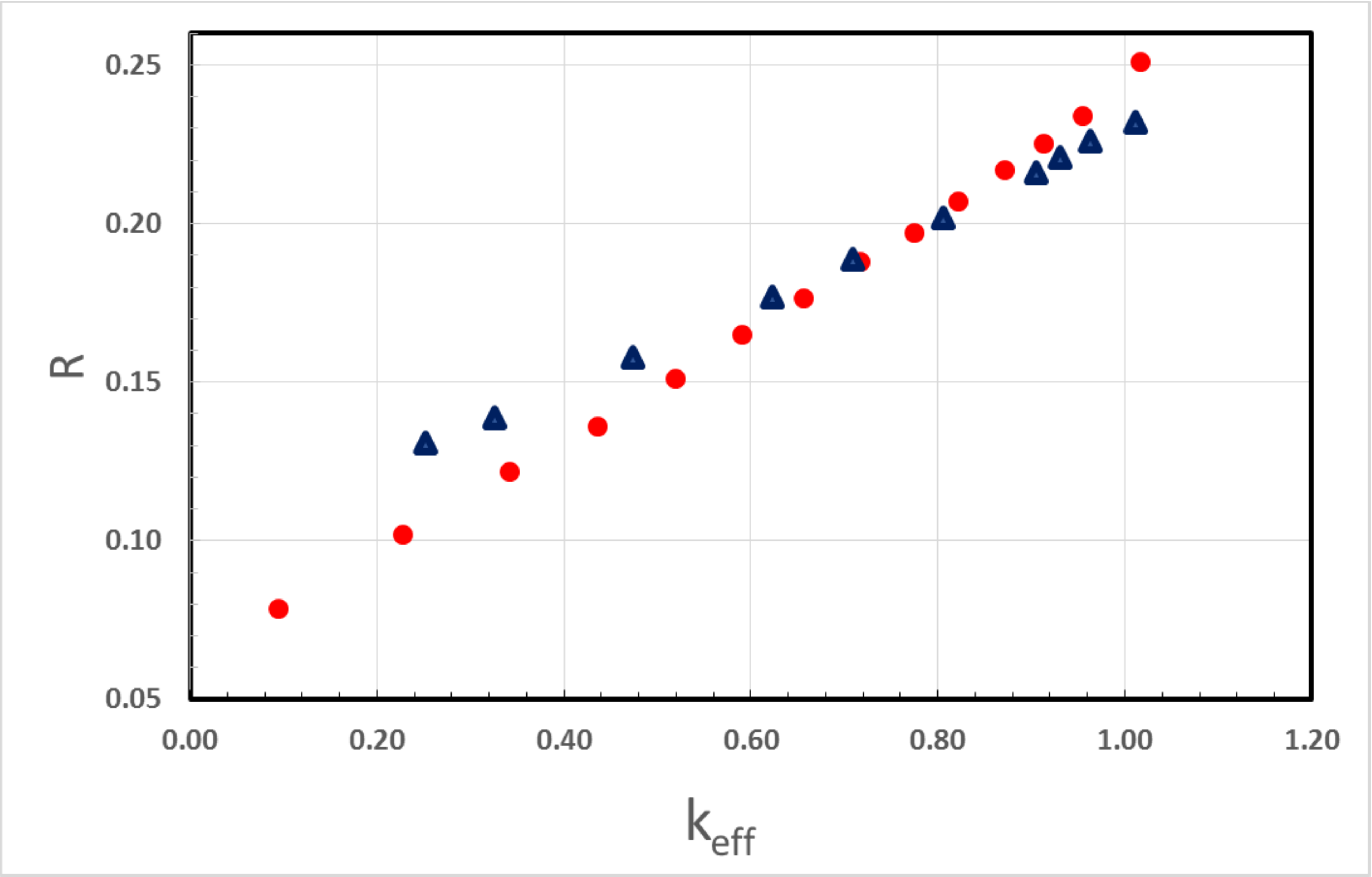}
\caption{R for model (4), a sodium cooled system (blue triangles), compared to model (1) (red circles). Error bars are smaller or equal to the marker size.}
\label{Fig:fig5}
\end{figure}

A separate analysis is required for sodium reactors (model (4)) where a slightly softer neutron spectrum may be expected due to the lighter coolant. Simulation results reported in Fig.~\ref{Fig:fig5} still show a smooth R($k_{\rm eff} )$, but both the slope and the critical R(1) value are  smaller compared with fast lead reactors, consistently with a   softer  energy spectrum. This result does not exclude some kind of convergence to be at work also for this kind of fast reactors, with the consequent possibility to use R as an observable to estimate $k_{\rm eff}$ in this case, too.

All the above results show that R, although not a universal function, appears to be relatively independent from the detailed structure of the core.

\section{Conclusions }

In conclusion, the previous discussion of our simulation results shows that for $k_{\rm eff}>0.6-0.7$ 
\begin{itemize}
\item for different models of fast subcritical reactors the ratio R approaches an almost linear function of $k_{\rm eff}$;
\item the functions for the various reactor models are quite similar, with slight differences in slope and approach to the critical limit, indicating a kind of convergence between different types of core;
\item the ratio R can be used as an observable for estimating $k_{\rm eff}$ once the R ($k_{\rm eff} )$ curve has been computed for a realistic reactor model and checked against experimental data on R in the reference critical configuration and against experimental data on R for various subcritical configurations, together with $k_{\rm eff}$ values obtained by other methods (which implies some renormalization or calibration procedure where for instance the whole curve may be shifted horizontally to match the experimental data).
\end{itemize}

Assuming the trends to be linear at medium/high $k_{\rm eff}$ leads to the following relative error:

\begin{equation}
\frac{\delta k_{\rm eff}}{ k_{\rm eff}}=\frac{\delta R}{ R}
\label{eqn:error}
\end{equation}
i.e., in the case of $^{235}$U-based fuel, an accuracy of 1\$ would require a 0.7\% error in the measurement of R.
Being  the measurement of R relative, the available precision should be mostly limited by the knowledge of the two-group efficiencies of the detectors: for instance using $^{235}$U and $^{238}$U fission chambers or, even better, the recently developed diamond/$^6$Li sandwich spectrometers \citep{diamond}, accuracies of 1\$ or less should be reachable in a time scale of the order of minutes to few hours for neutron fluxes of the order of $10^9-10^{10}$ n$/$cm$^2/$s.

Concerning systematic errors, one possibility is to consider them on a point-by-point basis, i.e. by taking the deviation of single points from the linear trends used for 
$k_{\rm eff}>$0.7. In the case of model(1) (which has the smallest statistical errors), the absolute deviation of a single point from the linear trend is at worst 0.0025, i.e. about 1 \% for the points considered in the linear fit. By assuming the linear behavior to be valid to first order,  Eqn.~\ref{eqn:error} gives a 1 \% error also on $k_{\rm eff}$, i.e. about 1.4 \$.

Concerning \Red{instead} systematic errors associated to the lack of a fully realistic description of the core, an initial comparison with experimental data on R in a start-up critical configuration and on both R and $k_{\rm eff}$  (the latter obtained by other methods) in various subcritical configurations, should practically make such systematic errors negligible. 
A similar consideration can be done for systematic $\Delta k^{\rm sys}_ {\rm eff}$ associated to uncertainties in the used nuclear data libraries. It is reasonable to expect that such uncertainties may cause, for high $k_{\rm eff}$ values, a shift of the computed R($k_{\rm eff})$ curve along the horizontal axis. 

Overall, all the above systematic effects may be compensated by renormalizing/shifting the whole curve, based on the measured R(1) value in the reference critical configuration and on the R and $k_{\rm eff}$ (by other methods) measurements performed in various subcritical configurations.

\Red{Moreover effects associated with any local change in the core materials may modify the nearby measured value of the ratio $R$ without
significantly affecting $k_{\rm eff}$`; they should be accounted particularly in accident conditions and should be investigated.}

Therefore, we conclude that the R observable, once properly calibrated, may be a valid, source- and beam current-independent additional tool for the online measurement of the reactivity in subcritical assemblies in all operational conditions. 

\section*{Acknowledgements}
This  work  is  partially  supported  by  the  7th  Framework  Programs  of  the  European 
Commission  (EURATOM)  through  the FREYA contract \# 269665.


\end{document}